\documentclass[reprint,prl,aps,superscriptaddress,longbibliography,preprintnumbers]{revtex4-2}
\usepackage[utf8x]{inputenc}
\usepackage{amssymb}
\usepackage{amsmath}
\usepackage{graphicx}
\usepackage{bbm}
\usepackage{psfrag}
\usepackage{latexsym}
\usepackage{color}
\usepackage[dvipsnames]{xcolor}
\usepackage{hyperref}
\hypersetup{colorlinks=true, citecolor=blue, urlcolor=blue, linkcolor=blue,urlcolor=cyan}
\usepackage{tensor}

\usepackage{amsfonts}
\usepackage{amsmath}
\allowdisplaybreaks[4]        
\usepackage{amssymb}
\usepackage{euscript}     
\usepackage{color}  
\usepackage{xcolor}       
\usepackage{tensor}        
\usepackage{amsthm}
\usepackage{graphicx}
\usepackage{subfigure}
\usepackage{bbm}
\usepackage[header,title,page,titletoc]{appendix}

\renewcommand{\(}{\left(}
\renewcommand{\)}{\right)}
\renewcommand{\[}{\left[}
\renewcommand{\]}{\right]}

\newcommand{\ket}[1]{{\left\vert{#1}\right\rangle}}

\newcommand{\nsb}{n}  

\newcommand{\be}{\begin{equation}}
\newcommand{\ee}{\end{equation}}
\newcommand{\bea}{\begin{eqnarray}}
\newcommand{\eea}{\end{eqnarray}}
\newcommand{\beq}{\begin{equation}}
\newcommand{\eeq}{\end{equation}}
\newcommand{\beqa}{\begin{eqnarray}}
\newcommand{\eeqa}{\end{eqnarray}}
\newcommand{\beqar}{\begin{eqnarray*}}
	\newcommand{\eeqar}{\end{eqnarray*}}

\newcommand{\eg}{{\it e.g.,}\ }
\newcommand{\ie}{{\it i.e.,}\ }

\newcommand{\reef}[1]{(\ref{#1})}

\newcommand{\mt}[1]{\textrm{\tiny #1}}

\newcommand{\mC}{\mathcal{C}}

\newcommand{\mL}{\mathcal{L}}

\newcommand{\GN}{G_\mt{N}}

\begin{document}

\author{Alexandre Belin}\email{a.belin@cern.ch}
\affiliation{\it CERN, Theory Division, 1 Esplanade des Particules, Geneva 23, CH-1211, Switzerland}

\preprint{CERN-TH-2021-181; YITP-22-02}

\author{Robert C. Myers}\email{rmyers@perimeterinstitute.ca}
\affiliation{\it Perimeter Institute for Theoretical Physics, Waterloo, Ontario N2L 2Y5, Canada}

\author{Shan-Ming Ruan}\email{sruan@perimeterinstitute.ca}
\affiliation{\it Perimeter Institute for Theoretical Physics, Waterloo, Ontario N2L 2Y5, Canada}
\affiliation{\it Department of Physics $\&$ Astronomy, University of Waterloo, Waterloo, ON N2L 3G1, Canada}
\affiliation{\it Yukawa Institute for Theoretical Physics, Kyoto University, Kitashirakawa Oiwakecho, Sakyo-ku, Kyoto 606-8502, Japan}

\author{G\'abor S\'arosi}\email{gabor.sarosi@cern.ch}
\affiliation{\it CERN, Theory Division, 1 Esplanade des Particules, Geneva 23, CH-1211, Switzerland}

\author{Antony J. Speranza}\email{asperanz@gmail.com}
\affiliation{\it Perimeter Institute for Theoretical Physics, Waterloo, Ontario N2L 2Y5, Canada}
\affiliation{\it Department of Physics, University of Illinois, Urbana-Champaign, Urbana IL 61801, USA}

\title{Does Complexity Equal Anything?}

\begin{abstract}
We present a new infinite class of gravitational observables in asymptotically Anti-de Sitter space living on codimension-one slices of the geometry, the most famous of which is the volume of the maximal slice. We show that these observables display universal features for the thermofield-double state: they grow linearly in time at late times and reproduce the switch-back effect in shock wave geometries. We argue that any member of this class of observables is an equally viable candidate as the extremal volume for a gravitational dual of complexity.
\end{abstract}

\maketitle

\noindent \emph{1.~Introduction:} Complexity theory aims to quantify how difficult it is to perform a chosen task using a set of simple operations. In quantum complexity theory, one implements the desired operation by constructing a quantum circuit with simple unitary gates. An important aspect to keep in mind is that in complexity theory, the interest is always on the scaling of the complexity with the ``size" of the problem (\eg the dimensionality of the Hilbert space in quantum complexity). 
Extracting a precise result for the complexity is often too complicated, and this number is expected to be highly sensitive, \eg to the choice of allowed simple operations, while the scaling is a robust property of the problem class.

Quantum complexity has recently triggered much interest in the context of black holes and holography, as a new twist in the ongoing effort connecting quantum information theory to quantum gravity. The length of the wormhole for a two-sided AdS black hole grows linearly in time at late times, and continues growing far beyond times at which entanglement entropies have thermalized \cite{Hartman:2013qma}. This suggests that a new quantum information measure is needed to encode the growth of the wormhole, and several holographic proposals were made relating it to quantum complexity, \eg complexity has been proposed to be dual to the volume of the maximal slice (CV proposal) \cite{Susskind:2014rva}, the action of the Wheeler-de Witt patch (CA proposal) \cite{Brown:2015bva} or the spacetime volume of the Wheeler-de Witt patch (CV2.0 proposal) \cite{Couch:2016exn}.

It is important to emphasize that the formulation of all three of the above proposals is ambiguous. For CV and CV2.0, this comes from an additional length scale needed to obtain a dimensionless number out of the volume. For CA, it results from ambiguities in the boundary terms on null slices \cite{Lehner:2016vdi}. However, rather than a shortcoming, this ambiguity can be seen as a feature of holographic complexity, as it connects nicely with the ambiguities arising in complexity theory, \eg the choice of a gate set. Therefore, a gravitational dual for complexity should reflect these conventional ambiguities.

 In this letter, we explore this idea and show that in fact, there is an infinite class of gravitational observables defined in a diffeomorphism-invariant way that display universal features and hence are equally viable candidates for a gravitational dual of complexity as the
volume or the action. These observables are defined on codimension-one regions of the geometry as
\begin{equation}\label{eq:obsdef}
O_{F_1,\Sigma_{F_2}}(\Sigma_{\textrm{CFT}}) =\frac{1}{\GN L} \int_{\substack{\Sigma_{F_2}  }}\!\!\!\!\! d^d\sigma \,\sqrt{h} \,F_1(g_{
	\mu\nu}; X^{\mu}) \,.
\end{equation}
where $F_1$ is a scalar function of the background metric $g_{\mu\nu}$ and of an embedding $X^\mu(\sigma^a)$ of the codimension-one surface $\Sigma_{F_2}$. Asymptotically $\Sigma_{F_2}$ is fixed by the boundary condition $\partial\Sigma_{F_2}=\Sigma_{\textrm{CFT}}$. In this letter, we will focus on the case where $\Sigma_{\textrm{CFT}}$ is a constant time slice in the boundary CFT, and so we often write $O_{F_1,\Sigma_{F_2}}(\tau)$ ($\tau$ is the CFT time). For such quantities to be diffeomorphism invariant, the bulk slice must be defined in a coordinate independent way. We determine this slice by requiring that it extremizes a particular scalar functional $F_2$, \ie
\begin{equation}
\delta_{\mt{X}} \left( \int_{\Sigma} d^d\sigma \,\sqrt{h} \,F_2(g_{\mu\nu};X^\mu)\right) =0 \,.
\end{equation}
Note that generally the scalar functions $F_1$ and $F_2$ need not coincide. For $F_1=F_2=1$, this prescription yields the extremal volume appearing in the CV proposal. Allowing for more general functions gives an infinite family of new observables. 

A class of such generalized functionals have appeared previously in the context of 
holographic complexity for higher curvature theories of gravity \cite{Alishahiha2015, Bueno2016,
Hernandez:2020nem}, in which the volume functional is corrected with higher curvature contributions.
The present work emphasizes that such functionals also serve as good measures of 
holographic complexity in standard general relativity
(or potentially any diffeomorphism-invariant gravitational theory), since they are examples of
observables $O_{F_1,\Sigma_{F_2}}$ with $F_1=F_2$.  

\vspace{4 pt}
\noindent \emph{1.~Summary of Results:} We show that for a large subset of the class of functionals $O_{F_1,\Sigma_{F_2}}$, the two following universal properties hold when probing the thermofield double state:
\begin{enumerate}
\item Observables grow linearly with time at late times
\begin{equation}\label{eq:linear}
\lim\limits_{\tau\to\infty}O_{F_1,\Sigma_{F_2}}(\tau) \sim P_{\infty}\, \tau  \,.
\end{equation}
In the limit of large temperature, the constant $P_{\infty}$ is proportional to the mass.

\item Observables exhibit the {\it switchback effect} \cite{Stanford:2014jda}, a universal time delay in response to perturbations described by particles falling into the dual black hole.
\end{enumerate}
 Properties 1 and 2 are expected to be displayed universally by any definition of quantum complexity and are held as the main evidence for the CV and CA proposals. The linear growth property is a consequence of viewing the time evolution operator $\exp(-i H \tau)$ as a quantum circuit, whose size scales as $\propto \tau$. On the other hand, the switchback effect is an imprint of the butterfly effect on complexity growth. We can insert an operator at some time $t$ in the past which corresponds to acting with $\exp(-iHt)W\exp(i Ht)$ on our state. If we don't add any operator $W$, the backward and forward time evolution exactly cancel. With the operator, the cancellation is disturbed but still occurs approximately for some time, before the effect of the operator grows large due to the butterfly effect. This logic suggests the size of the circuit implementing $\exp(-iHt)W\exp(i Ht)$ for large $t$ is $\sim 2P_{\infty}\,(t-t_*)$, where $t_*$ is the scrambling time \cite{Stanford:2014jda}. 

The universality displayed by the class of observables \eqref{eq:obsdef} leads us to conclude that any of them are equally good candidates for a gravitational dual of complexity. This nicely parallels  the expected ambiguities for quantum complexity, noted above, where a precise definition of the complexity depends on many choices. Hence the precise value of the complexity is unimportant, whereas the scaling (here, the time dependence) is universal.

\begin{figure}[ht!]
	\centering\includegraphics[height=2.5in]{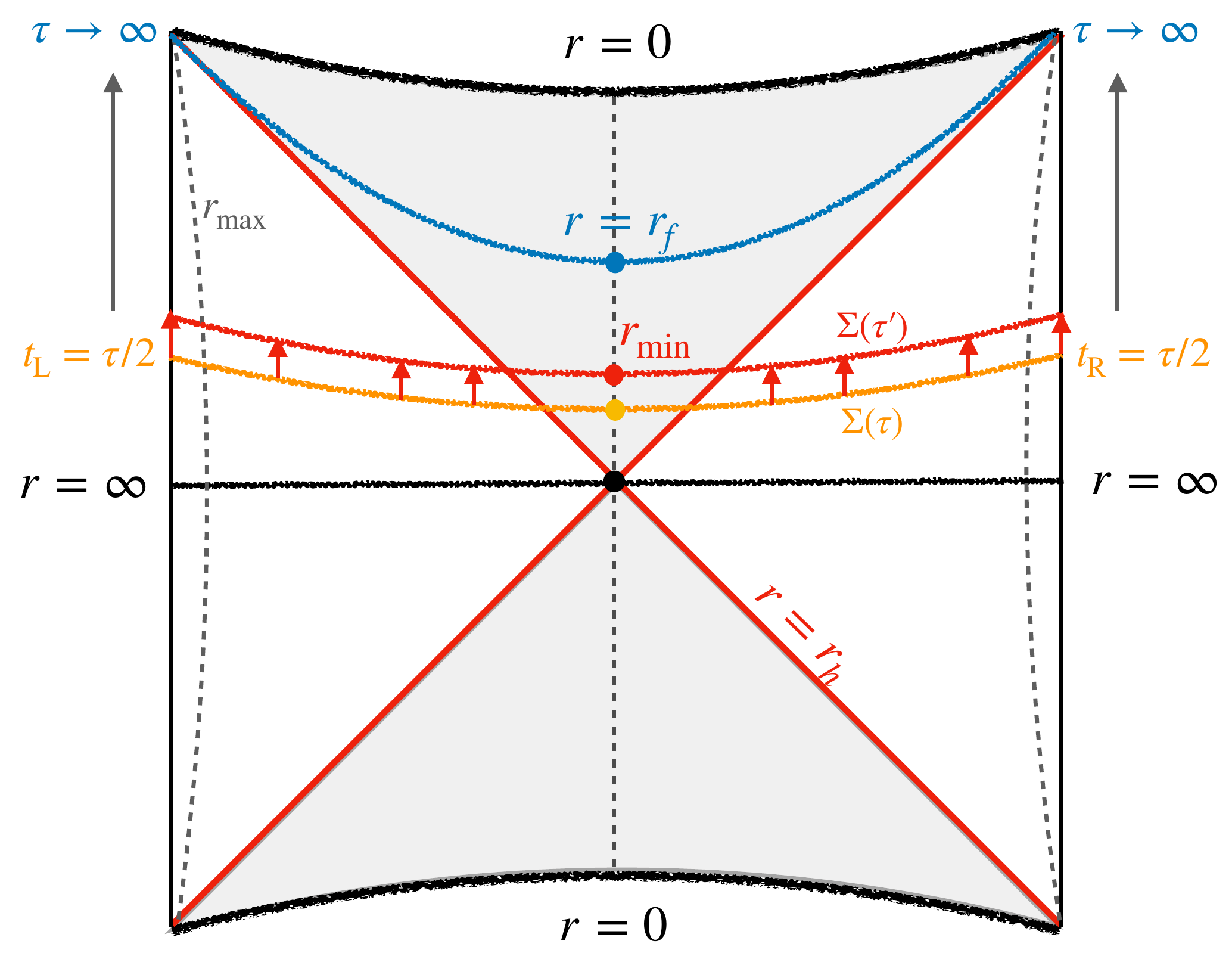}
	\caption{The time evolution of the extremal hypersurfaces from \textcolor{orange}{$\Sigma(\tau)$} to a nearby extremal hypersurface \textcolor{red}{$\Sigma(\tau')$}. At infinite time limit $\tau \to \infty$, the extremal surface approaches a constant-$r$ hypersurface at \textcolor{blue}{$r=r_f$} where the effective potential arrives at a local maximal value.}\label{fig:time}
\end{figure}

\vspace{4 pt}
\noindent \emph{2.~Observables with $F_1=F_2$:}  
The quantities \eqref{eq:obsdef} provide a huge class of diff-invariant observables in AdS which can probe a plethora of states, ranging from small perturbations of AdS to black holes. In this letter, we focus on the eternal planar black hole, which corresponds to two decoupled CFTs on planar (\ie $\mathbb{R}^{d-1}$) spatial slices entangled in the thermofield double (TFD) state 
\begin{equation}\label{eq:TFD}
\ket{\psi_{\mt{TFD}}(\tau)}=\sum_{E_n} e^{-\beta E_n/2-iE_n \tau } \ket{n}_{\mt{L}} \otimes \ket{n}_{\mt{R}} \,.
\end{equation}
The spacetime dual to this state is described 
in Eddington-Finkelstein coordinates by the metric
\begin{equation}\label{eq:vmetricBH}
d s^{2}=-f(r) d v^{2}+2\,dv\,dr+\frac{r^{2}}{L^2}\, d{\vec x}^{\,2}\,,
\end{equation}
where $f(r)= \frac{r^2}{L^2}\big( 1 - \frac{r_h^{d}}{ r^{d}}\big)$
and we use the infalling coordinate $v=t + r_*(r)$ with $r_*(r)= -\int^\infty_r \frac{dr'}{f(r')}$ to cover the black hole interior. The spacetime describes the time evolution of the CFT state living at time $t_{\mt{R}}=t_{\mt{L}}=\tau/2$ (see Fig.~\ref{fig:time}).

We will start by considering observables with $F_1=F_2$, such that the slice on which the functional is evaluated extremizes the functional itself. We will call these observables $\mathcal{C}_{\textrm{gen}}$, as they generalize the CV proposal. To probe the state \eqref{eq:TFD}, we must anchor the surface on the CFT slice $\Sigma_\tau$, \ie the boundary of the hypersurface $\Sigma(\tau)$. Thus 
\begin{equation}\label{eq:dCdt}
\mC_{\rm{gen}}(\tau) =\max_{\partial\Sigma(\tau)=\Sigma_\tau} \[ \frac{1}{G_{\mt{N}} \,L}  \int_{\Sigma} \! d^d\sigma \,\sqrt{h} \,F_1(g_{\mu\nu};X^\mu(\sigma))  \].
\end{equation}

Due to the symmetries of the state and the hypersurface $\Sigma(\tau)$, we can parametrize our surface with a single function $r(v)$. When $F_1=F_2$, finding the extremal surface is analogous to solving a classical mechanics problem where the action and Lagrangian are identified as $S\sim \mC_{\rm{gen}}$, $\mathcal{L} \sim F_1\sqrt{h}$. Moreover, the extremality conditions become the equations of motion derived from $\mL$.

In this picture, the time derivative of $\mathcal{C}_{\textrm{gen}}$ evaluated on the extremal surface is related to the momentum at the endpoint of the trajectory: 
\begin{equation}\label{eq:dCdtR}
\frac{d\mC_{\rm{gen}}}{d\tau} = \frac{1}{2} \,\mathcal{P}_t\bigg|_{\partial \Sigma(\tau)} \equiv  \frac{1}{2}\,\frac{\partial \mathcal{L}}{\partial \dot t}\bigg|_{\partial \Sigma(\tau)} \,,
\end{equation}
where $\mathcal{P}_t$ is the momentum conjugate to the coordinate time $t$. 
Linear growth amounts to the statement that $\mathcal{P}_t|_{\partial \Sigma(\tau)}$ approaches a constant at large $\tau$.

\vspace{4 pt}
\noindent \emph{3.~Time evolution of the extremal surfaces:} 
We will now proceed to solve for the location of the extremal surfaces as a function of time and study the late time behavior of our new observables.
Due to the symmetries of the planar black hole, we can parametrize the spacelike hypersurface $\Sigma$ simply by $(v(\sigma), r(\sigma),\vec{x})$. Derivatives with respect to $\sigma$ will be represented by dots. In this letter, we will focus on observables where the function $F_1(g_{\mu\nu}, \mathcal{R}_{\mu\nu\rho\sigma}, \nabla_\mu)$ depends only on $(d+1)$-dimensional curvature invariants evaluated on the extremal slice $\Sigma$. (We comment on more general observables in the discussion.) For such observables, one can rewrite the generalized volume in our parametrization by
\begin{equation}\label{eq:defineV}
\mC_{\rm{gen}}=\frac{V_x }{\GN L} \int_\Sigma d\sigma\,\(\frac{r}{L}\)^{d-1}\!\sqrt{-f(r\,){\dot v}^2+2\dot v\,\dot r}\ a(r)\,,
\end{equation}
where $\sigma$ can be understood as a radial coordinate on the hypersurface $\Sigma$, $V_x$ denotes the (regulated) volume of the spatial boundary directions $\vec x$, and the factor $a(r)$ is the result of evaluating the corresponding function of curvatures on the surface. The restriction to $(d+1)$-dimensional curvature invariants was made so that the factor $a(r)$ does not depend on derivatives of $r$, which considerably simplifies the problem \footnote{See Supplemental Material (appendix A) for a detailed analysis involving a Weyl-squared term $C_{\mu\nu\rho\sigma}C^{\mu\nu\rho\sigma}$.}.

\begin{figure}[ht!]
	\centering\includegraphics[height=2.7in]{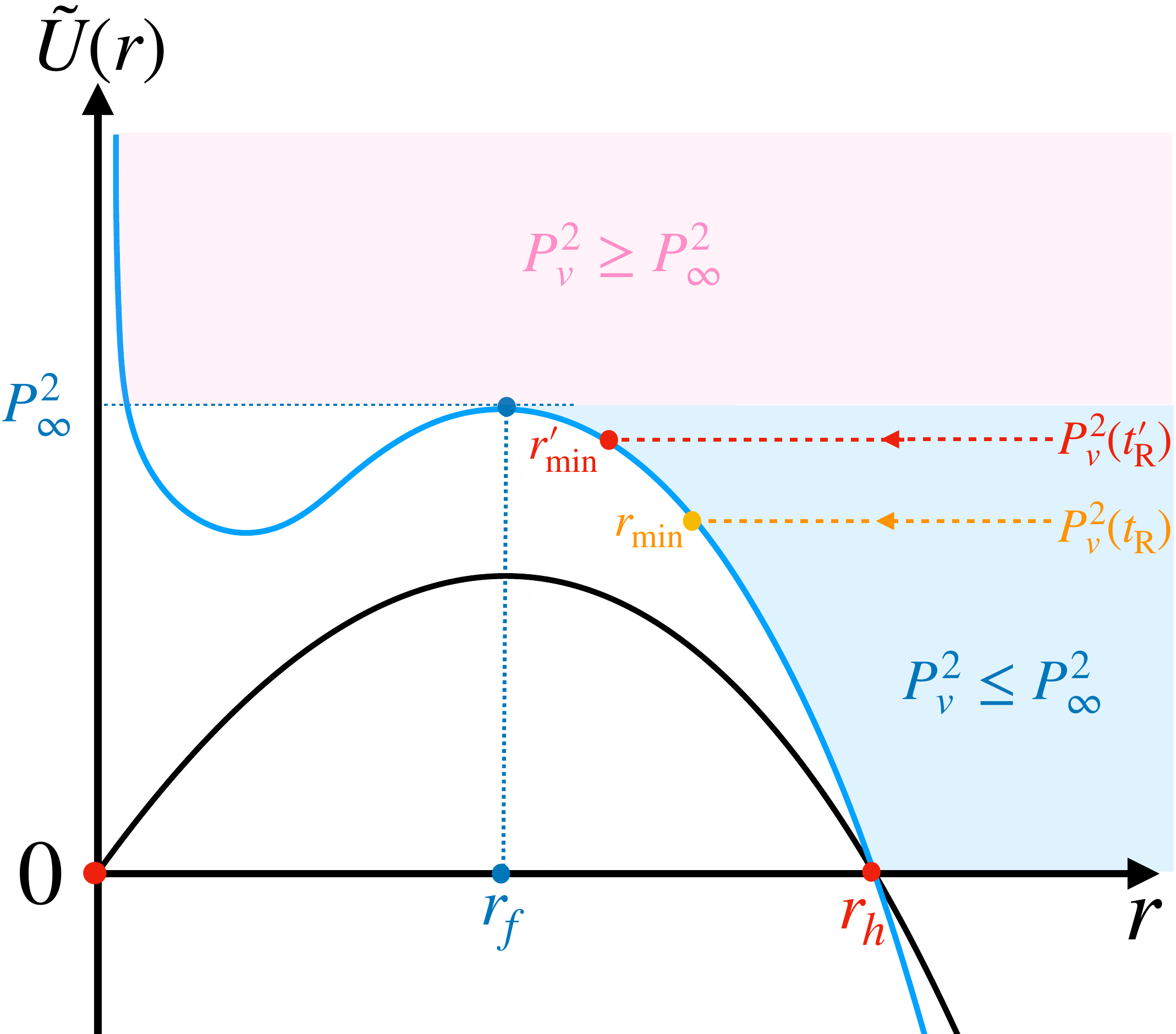}
	\caption{The blue curve denotes a characteristic effective potential $\widetilde{U}(r)$ depending on the spacetime curvature and the black curve presents the potential from the volume. The turning point at the minimal radius $r_{\rm{min}}$ satisfies $\dot{r}=0$ which is equivalent to $P_v^2=\widetilde{U} (r_{\rm{min}})$ for a given conserved momentum. The critical value of the momentum $P_{\infty}$ is obtained at the infinite boundary time $\tau \to \infty$.}\label{fig:potential}
\end{figure}

Of course, we also require that the functionals $\mC_{\rm{gen}}$ are diffeomorphism invariant. Consequently, one can easily show that $\mC_{\rm{gen}}$ is invariant under the transformation $\sigma \to g(\sigma)$. As a result, we can fix the gauge by choosing
\begin{equation}\label{eq:gauge}
\sqrt{-f(r\,){\dot v}^2+2\dot v\,\dot r} = 
a(r)\(\frac{r}{L}\)^{d-1}\,.
\end{equation}
Because the bulk spacetime is stationary, one finds that the momentum $P_v$ conjugate to the infalling time $v$, \ie
\begin{equation}\label{eq:vmoment}
P_v
=\frac{a(r)\,(r/L)^{d-1}\left(\dot r -f(r)\,\dot v\right)}{\sqrt{-f(r\,){\dot v}^2+2\dot v\,\dot r}}=\dot r -f(r)\,\dot v\,,
\end{equation}
is conserved on the extremal surfaces. Combining these two equations, one arrives at the extremality conditions:
\begin{equation}\label{eq:dots}
\begin{split}
\dot r&=\pm\sqrt{P_v^2+f(r)\,a^2(r)\,\(\frac{r}{L}\)^{2(d-1)}}\,,
\\
\dot v&=\frac{1}{f(r)}\left(
-P_v\pm\sqrt{P_v^2+f(r)\,a^2(r)\(\frac{r}{L}\)^{2(d-1)}}\right)\,.
\end{split}
\end{equation}
We can recast this problem as the motion of a classical non-relativistic particle in a potential \cite{Carmi:2017jqz,Chapman:2018lsv}. To wit, 
\begin{equation}\label{eq:Hamiltonian}
\dot{r}^2 +\widetilde{U}(r) = P_v^2  \ \ \ \text{with} \ \ \ \widetilde{U}(r)= -f(r) a^2(r) \(\frac{r}{L}\)^{2(d-1)} \!\!,
\end{equation}
with an effective potential $\widetilde{U}$ which parametrizes the choice of observable. Fig. \ref{fig:potential} presents a characteristic potential. Because of the multiplicative prefactor $f(r)$, the effective potential vanishes at the horizon  $r=r_h$. We will particularly be interested in the case where the effective potential has at least one local maximum inside the horizon. As we will show, this is precisely the requirement for the linear growth of the generalized volume. When the couplings for the higher curvature terms in $F_1$ are sufficiently small, such a local maximum always exists [10]. We can specify the local maximum at $r=r_f$ by 
\begin{equation}
\widetilde{U}(r_f)= P^2_{\infty}\,, \quad \widetilde{U}'(r_f)=0\,,\quad  \widetilde{U}''(r_f)\le 0 \,.
\end{equation}
Finding the extremal surface anchored on a specific boundary time slice $\tau$ thus corresponds to solving the Hamiltonian system \eqref{eq:Hamiltonian} with a given conserved momentum $P_v$ that is fixed by the boundary time, namely \footnote{We note that the integrand is divergent at the horizon because $f(r) \sim f'(r_h)(r-r_h)$. Hence this integral \reef{eq:boundarytimetR02} is defined by the Cauchy principal value associated with this singularity, which is finite.}
\begin{equation}\label{eq:boundarytimetR02}
\tau\equiv 2t_{\mt{R}}= -2\int^{\infty}_{r_{\rm{min}}} dr\, \frac{P_v}{f(r)\sqrt{P_v^2-\widetilde{U}(r)}}  \,.
\end{equation}
We denote $r_{\rm{min}}$ as the minimal radius lying on the timelike surface $t=0$ (dashed vertical line in Fig.~\ref{fig:time}) and determined by the conserved momentum through $P_v^2=\widetilde{U}(r_{\rm{min}})$. This is also the turning point of the analogue particle. As discussed in eq.~\eqref{eq:dCdtR}, the time derivative of the generalized volume at boundary time $\tau$ is given by
\begin{equation}\label{eq:dVdt01}
\frac{d \mathcal{C}_{\rm gen}}{d\tau} =\frac{1}{2}\,\mathcal{P}_t\big|_{\partial \Sigma}=\frac{V_x }{\GN L}\,P_v(\tau)\,.
\end{equation}

\noindent \emph{4.~Linear Growth at Late Times:\label{sec:linear}} For any boundary time $\tau$, fixing the conserved momentum $P_v(\tau)$ via eq.~\eqref{eq:boundarytimetR02}, there is a corresponding extremal hypersurface. For example, $P_v^2=0$ gives the time reflection symmetric slice $t=0$. For generic $\tau$, the radial trajectory of the extremal surface starts from the asymptotic boundary and moves into the interior of the black hole until it reaches the minimal radius $r_{\rm{min}}$. Increasing the conserved momentum $P_v$ to the critical value $P_{\infty}$ causes the boundary time to diverge because the extremal surface is near the final slice at $r=r_f$. This can be seen by expanding the potential around the maximum
\begin{equation}
P_{\infty}^2-\widetilde{U}(r)  \sim - \frac{1}{2} \widetilde{U}''(r_f)(r-r_f)^2+\mathcal{O}((r-r_f)^3) \,.
\end{equation}
and substituting into \eqref{eq:boundarytimetR02}. Then we note that a divergence arises from the lower boundary of the integral as $r_{\rm min}$ approaches $r_f$.
Similarly, we will now demonstrate that $\lim_{\tau \rightarrow \infty} P_v(\tau)= P_{\infty}$. Due to \eqref{eq:dVdt01}, this means that the generalized volume grows linearly at late times.
Before doing so, we stress again that the existence of the above late-time limit requires the existence of a local maximum of the effective potential inside the horizon. 

Using eq.~\eqref{eq:boundarytimetR02}, it is straightforward to get
\begin{equation}\label{eq:dtRdPv}
\begin{split}
\frac{d \tau}{d P_v} &= \frac{d r_{\rm{min}}}{d P_v}\, \frac{2P_v}{ f(r) \sqrt{P_v^2 - \widetilde{U}(r)}} \Bigg|_{r\to r_{\rm{min}}}  \\
&\quad+ \int^{\infty}_{r_{\rm{min}}} dr\, \frac{2\widetilde{U}(r)}{f(r) \(P_v^2-\widetilde{U}(r)\)^{3/2}}\,,
\end{split}
\end{equation}
where both terms on the right hand side are divergent due to $P^2_v - \widetilde{U}(r_{\rm{min}})=0$. Approaching the infinite boundary time by pushing the minimal radius to the final slice (\ie $r_{\rm{min}} \to r_f$ or  equivalently $P_v \to P_{\infty}$), the leading divergence appearing in eq.~\eqref{eq:dtRdPv} in the late-time limit is 
\begin{equation}\label{eq:dtaudPv}
\frac{d \tau}{d P_v}\sim  \frac{-2\sqrt{2}\,\widetilde{U}(r_f)}{f(r_f)  \( - \widetilde{U}''(r_f)   \)^{\frac{3}{2}}} \frac{1}{\(r_{\rm{min}}-r_f\)^2}\,. 
\end{equation}
More explicitly, the above limit implies that in the late-time limit, the growth rate $P_v (\tau)$  approaches the critical value $P_{\infty}$ from below as an exponential
\begin{equation}\label{eq:Pvdecay}
P_{\infty} - P_v(\tau)   \propto e^{- \kappa \tau}\,, \text{with}\quad \kappa = \frac{-f(r_f)\sqrt{-2\widetilde{U}''(r_f)}}{P_{\infty}}\,.
\end{equation}
Note that $\kappa$ is always positive, since the final slice is inside the horizon (hence $f(r_f) < 0$) and corresponds to a local maximum of the potential with $\widetilde{U}''(r_f) <0$.

Finally, let us show that $P_{\infty}\propto M$ where $M\propto \frac{r_h^d}{\GN L^2}$ is the ADM mass of the black brane. We can introduce a dimensionless radial coordinate $w = (r/r_h)^d$, in terms of which the potential is rewritten as
\begin{equation}\label{eq:nondimpotential}
\widetilde U(r)=\( \frac{r_h}{L} \)^{2d}  \(  w-w^2   \) a^2 (w)\,.
\end{equation}
We note that $a(w)$ depends only on $w$ because curvature invariants in the geometry \eqref{eq:vmetricBH} are independent of $r_h$ when written as a function of $w$. This is easy to see by using rescaled coordinates $\frac{r_h}{L}v,\frac{r_h}{L}\vec{x}$, in which metric components have no dependence on $r_h$. The dimensionful coefficients appearing in the definition of $F_1$ should not depend on $r_h$ either, because we want to define the observable in a state independent way. Using the recast potential  \eqref{eq:nondimpotential}, it is easy to show that its extrema satisfies
\begin{equation}
P_{\infty}^2\equiv \widetilde U(r_f) = \left( \frac{r_h}{L} \right)^{2d} \xi,
\end{equation}
where the constant $\xi$ is independent of $r_h$ \footnote{This result relies on having a scale invariant planar horizon, corresponding to a thermodynamic limit in the CFT. For compact horizons, there are finite size corrections, just like in the case of the CV proposal \cite{Carmi:2017jqz}.}.

\vspace{4 pt}
\noindent \emph{5.~More general observables with $F_1\ne F_2$:} We can also consider observables that are associated with the function $F_1$ but are evaluated on the extremal surface determined by the functional $F_2$. Let's focus on the simplest case where both $F_1$ and $F_2$ only depend on the bulk spacetime geometry. To wit, we define the observable as
\begin{equation}\label{eq:defineO12}
O_{F_1,\Sigma_{F_2} }\propto \int_{\Sigma_{F_2}} d\sigma\(\frac{r}{L}\)^{d-1}\sqrt{-f(r){\dot v}^2+2\dot v\,\dot r} \,a_1(r),
\end{equation}
where the extremal hypersurfaces $\Sigma_{F_2}$ are associated with the effective potential $\widetilde{U}_2(r)= -f(r) a_2^2(r) \(\frac{r}{L}\)^{2(d-1)}$. For $F_1=F_2$, the time derivative was given by the boundary term $P_v$. Now, the time derivative of this new observable with respect to the boundary time $\tau$ reads instead 
\begin{equation}
\begin{split}
&\frac{\GN L}{V_x }\,\frac{dO_{F_1,\Sigma_{F_2}}}{d\tau} = \sqrt{\bar{U}_1}\ + \\ 
& \quad 2 P_v \frac{d P_v}{d \tau} \int^{\infty}_{r_{\rm{min}}} dr\, \frac{ \sqrt{\widetilde{U}_1(r)\widetilde{U}_2(r)} -\sqrt{\frac{\bar{U}_1}{\bar{U}_2}}\widetilde{U}_2(r) }{f(r) \(P_v^2-\widetilde{U}_2(r)\)^{3/2}}\,,
\end{split}
\end{equation}
where we defined $\bar{U}_i = \bar{U}_i\(r_{\rm{min}}\)$ and the second term involving the integral vanishes when $\widetilde{U}_1=\widetilde{U}_2$. As with the analysis above, we find the new observable \eqref{eq:defineO12} still exhibits the expected linear growth at late times, \ie 
\begin{equation}
\begin{split}
\lim_{\tau \to \infty} \frac{dO_{F_1,\Sigma_{F_2}}}{d\tau} & =\frac{V_x }{\GN L}\sqrt{\widetilde{U}_1(r_f)}\,,\\
\end{split}
\end{equation}
where $r_f$ is the position of the maximum of $\widetilde{U}_2(r)$. But the growth rate is controlled by the functional $F_1$ \footnote{See Supplemental Material (appendix B) for more details.}.

\vspace{4 pt}
\noindent \emph{6.~Switchback Effect:} Consider the perturbed TFD state
\begin{equation}
\label{eq:switchbackstate}
|\Psi(t_{\mt{L}},t_\mt{R})\rangle = e^{-i H_{\mt{L}} t_{\mt{L}}-i H_{\mt{R}} t_{\mt{R}}}W_{\mt{L}}(t_n)...W_{\mt{L}}(t_1)|\psi_{\rm TFD}(0)\rangle,
\end{equation}
where the $t_1$,...,$t_n$ are in an alternating ``zig-zag" order. The corresponding complexity is then proportional to \cite{Stanford:2014jda} 
\begin{equation}
\label{eq:switchbackcomplexity}
 |t_\mt{R}+t_1|+|t_2-t_1|+\cdots +|t_\mt{L}-t_n|-2\nsb\, t_* \,.
\end{equation}
Here, the terms with absolute values give the total length of the path integral contour. 
The switchback effect is the subtraction $-2\nsb\, t_*$, owing to the partial cancellation of forward and backward parts of the time contours around the ``switchbacks", where the operator insertions have not been scrambled yet. The formula assumes that the time along each switchback (\ie each term involving an absolute value) is much larger than the scrambling time.

The state \eqref{eq:switchbackstate} is dual to a long wormhole supported by alternating left- and right-moving shockwaves. When all shocks are strong (which we study here), the geometry is obtained by gluing together AdS black hole geometries \eqref{eq:vmetricBH} along their horizons \cite{Shenker:2013yza}.
There are $n+1$ such patches, and the wormhole region consists of $n-1$ future and past interiors glued together in an alternating manner. The gluing is best described in Kruskal coordinates
\begin{equation}
\begin{split}
\label{eq:kruskalc}
U\cdot V&= - e^{f'(r_h) r_\ast (r)} \,,\quad \frac{U}{V} = - e^{-f'(r_h) t} \,, 
\end{split}
\end{equation}
in which the corresponding null shifts along the horizons are given by $\alpha_i= 2 e^{-\frac{2\pi}{\beta}(t_* \pm t_i)}$ with the inverse temperature $\beta = \frac{4\pi }{ f'(r_h)} $ and the scrambling time $t_*\propto \beta \log \GN^{-1}$.

The switchback effect associated with the maximal volume in this geometry was analyzed in \cite{Stanford:2014jda}. It is based on two properties of the extremal volume: (i) that it adds up in shockwave geometries, \ie
\begin{equation}\label{eq:piecewisevolume}
\begin{split}
\mathcal{V}&=\mathcal{V}(t_{\mt{R}},V_1)+\mathcal{V}(V_1+\alpha_1,U_2)+\cdots \\
\quad&+\mathcal{V}(U_{n-1}-\alpha_{n-1},V_n)+\mathcal{V}(V_n+\alpha_n,t_{\mt{L}}),
\end{split}
\end{equation}
where $\mathcal{V}(.,.)$ denotes the volume of the extremal slice connecting two specified points (either $V_i, U_i$ on a horizon or $t_{\mt{L,R}}$ on the boundary) in the black hole geometry; (ii) that $\mathcal{V}(.,.)$ all present a linear growth at late times. In order to find the maximal volume slice,  one needs to extremize the additive volume in \eqref{eq:piecewisevolume} with respect to $V_1,U_2,...U_{n-1},V_n$ which leads to the result  \eqref{eq:switchbackcomplexity}.
 
 We find that these two properties are also shared by the general observables $O_{F_1,\Sigma_{F_2}}$ defined in eq.\eqref{eq:defineO12}. The property (i) could in general fail when the functional involves curvature terms, since there could be extra $\delta$-function contributions at the shockwaves, spoiling the additivity property. However, we can prove that the diffeomorphism invariance of the functional $F_1$ guarantees that no contribution arises from $\delta$-functions on the shockwaves. Consider the geometry \eqref{eq:vmetricBH} in Kruskal coordinates 
 \begin{equation}
 d s^{2}=- 2A(UV) \,d U dV + B(UV)\, d{\vec x}^{\,2}\,,
 \end{equation}
 where $A= - \frac{2}{UV} \frac{f(r)}{f'(r_h)^2}, B= r^2/L^2$ are both functions of $UV$ because the spacetime is stationary. 
 For simplicity, we consider a single null shockwave at $U=0$. The backreaction on the geometry is described by 
 \cite{Sfetsos:1994xa,Shenker:2013yza,Stanford:2014jda}
 \begin{equation}
 \begin{split}
 \label{eq:shockwave}
 d s^{2}
 &=-2 A(U [V+\alpha \Theta( U)])d U d V + B(U  [V+\alpha \Theta(U)]) d{\vec x}^{\,2}\,,\\
 \end{split}
 \end{equation}
 where $\Theta( U)$ denotes the Heaviside step function. Suppose we want to write down a scalar function $F_1$ which only depends on the bulk geometry. All metric components depend on $U$ and $V$ as $f( U[V+\alpha \Theta(U)])$. Therefore, delta functions can come only from $U$ derivatives of metric components. Since $g_{UU}=g_{V V}=0$, in order to form scalars we must eventually contract all indices coming from derivatives with metric components $g^{U V}$, \eg $g^{UV}\partial_V\partial_U f$. As a result, one can show that all $U$ derivatives must come multiplied by an equal number of $V$ derivatives in scalar functions. Noting that $\partial_{ V}f(U[V+\alpha \Theta(U)])=U f'(U[V+\alpha \Theta(U)])$, we find that delta functions always come in the form $U \delta(U)$, $U^2 \delta'( U)$, etc. Since in the Kruskal geometry the functions $A$ and $B$ are regular and non-zero on the horizon we conclude that no delta functions appear in spacetime scalars formed from the shockwave geometry \footnote{Clearly, this result will also hold for more general stationary spacetimes, \eg black holes with spherical or hyperbolic horizons. One can view this as a generalization of a similar result \cite{Horowitz:1999gf} for AdS shockwaves to static Kruskal geometries.}.
 
Regarding the property (ii), we can show following section 4 that the general observables $O_{F_1,\Sigma_{F_2}}$ are still dominated by linear growth in ingoing/outgoing time, \ie 
 \begin{equation}
 \label{eq:volumegrowthformula}
O_{F_1,\Sigma_{F_2}} (U_{\mt{L}},V_{\mt{R}})= \frac{V_x }{\GN L}\sqrt{\widetilde{U}_1(r_f)}\, \big| v_{\mt{R}} +u_{\mt{L}} \big|+ O(1), 
 \end{equation}
 where $\widetilde{U}_1(r_f)$ denotes the value of the effective potential on the final slice and the the Kruskal coordinate is given by $|U|=e^{-\frac{2\pi}{\beta}u}$, $|V|=e^{\frac{2\pi}{\beta}v}$. Using the additive formula in \eqref{eq:piecewisevolume} for $ O_{F_1,\Sigma_{F_2}}$ and extremizing with respect to $V_1,U_2,...U_{n-1},V_n$, one recovers the expected switchback effect \eqref{eq:switchbackcomplexity}.
 


\vspace{4 pt}
\noindent \emph{7.~Discussion:} 
In this letter, we have introduced an infinite family of gravitational observables $ O_{F_1,\Sigma_{F_2}}$ defined on codimension-one slices of the geometry. We examined a large class of such observables, where the functionals $F_1,\,F_2$ only depend on 
spacetime curvature invariants, and showed they display universal behaviour: in the time-evolved thermofield double state, they grow linearly in time at late times and exhibit the switchback effect. However, we note that this universal behaviour also requires  the couplings on the curvature invariants not be too large. More general observables including functionals that depend independently on extrinsic and intrinsic data of the surface will be discussed in \cite{longpaper}, and we expect our results to extend to these cases as well. We now conclude with some open questions.

In light of the CA proposal, a natural question is to ask whether it is also possible to engineer codimension-0 observables like the action of the Wheeler-de Witt patch. In fact, it is possible to embed both codimension-0 and codimension-one observables in a unified framework following a construction by Peierls \cite{Peierls:1952cb}. This will be presented in \cite{longpaper}, where we will also discuss how to extract variations of these observables from the CFT, for coherent states of the gravitational theory \cite{Botta-Cantcheff:2015sav,Marolf:2017kvq,Belin:2018fxe,Belin:2018bpg,Belin:2020zjb,Bernamonti:2019zyy} using the dictionary between bulk and boundary symplectic forms. 

Finally, we have given strong evidence that a wide class of observables are all viable candidates for complexity. It would  be interesting to try and make the connection between ambiguities in the definition of quantum complexity, and the choice of gravitational observables more precise. However, is there perhaps a reason that would single-out the volume? The maximal volume not only captures the saturation of complexity at very late times \cite{Iliesiu:2021ari} but also satisfies nicer properties like existence theorems for the maximal slice \cite{witten2017}, positivity of the vacuum-subtracted volume \cite{Engelhardt:2021mju}, regular behaviour under small deformations \cite{Flory:2019dqx} and serving as an internal Hamiltonian of the Wheeler-de Witt patch \cite{Belin:2018bpg,Belin:2020zjb}. It would be interesting to explore this question further in the future.

\begin{acknowledgments}
\noindent \emph{Acknowledgments:}
We are happy to thank Jan de Boer, Shuwei Liu, Tadashi Takayanagi and Michael Walter for fruitful discussions and useful comments.  Research at Perimeter Institute is supported in part by the Government of Canada through the Department of Innovation, Science and Economic Development Canada and by the Province of Ontario through the Ministry of Colleges and Universities. RCM is supported in part by a Discovery Grant from the NSERC of Canada, and by the BMO Financial Group.  RCM also received funding from the Simons Foundation through the ``It from Qubit'' collaboration. AJS is supported by the Air Force Office of Scientific Research under award number FA9550-19-1-036.

\end{acknowledgments}

\bibliography{cvhighercurvature}


\pagebreak

\appendix

\pagebreak
\titlepage

\setcounter{page}{1}

\begin{center}
	{\LARGE Supplemental Material}
\end{center}

\section{A. Explicit example with a Weyl-squared term for $F_1 =F_2$}\label{app:A}

\begin{figure}[ht!]
	\centering
	\includegraphics[width=3.3in]{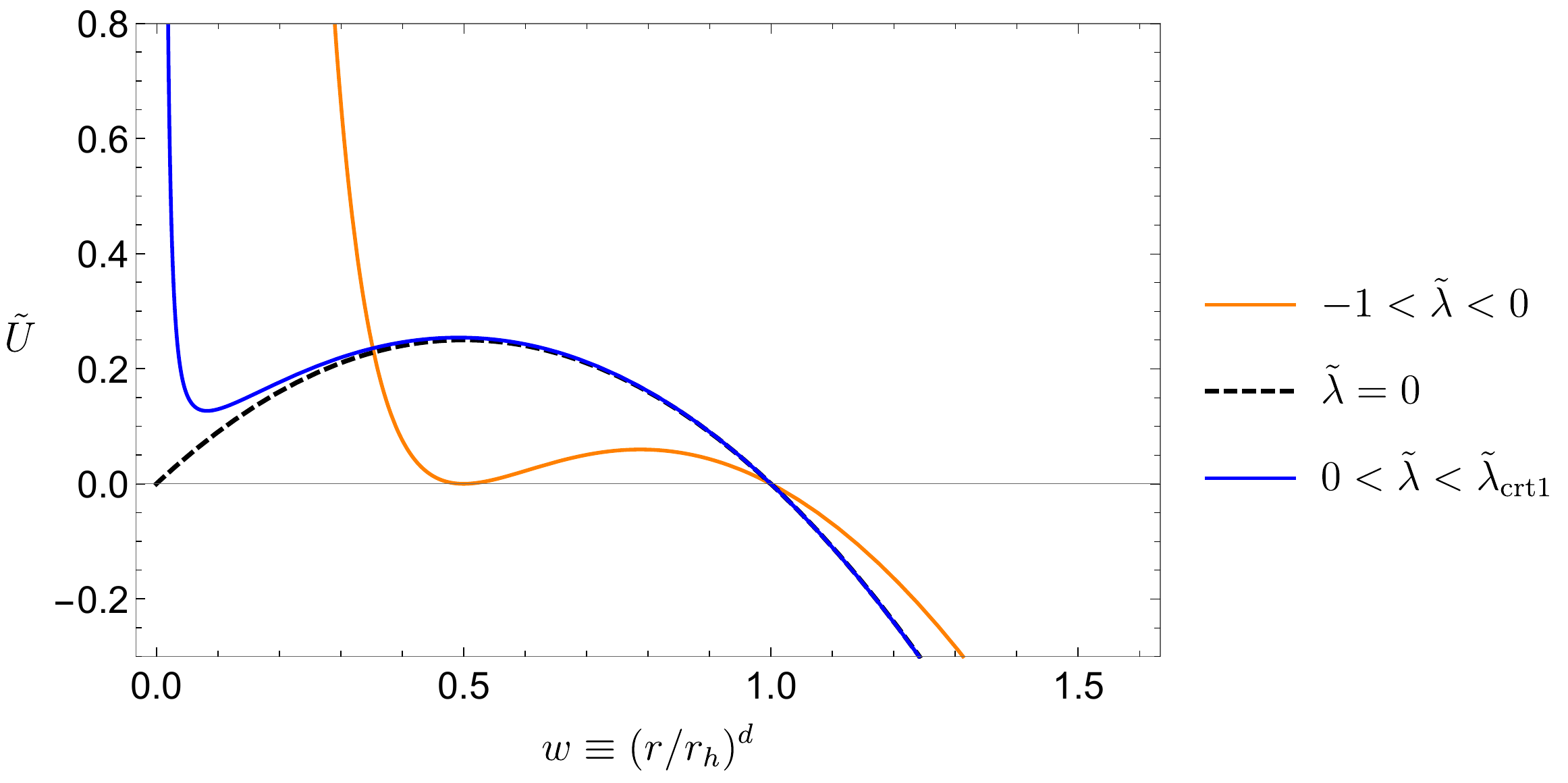}
	\includegraphics[width=3.5in]{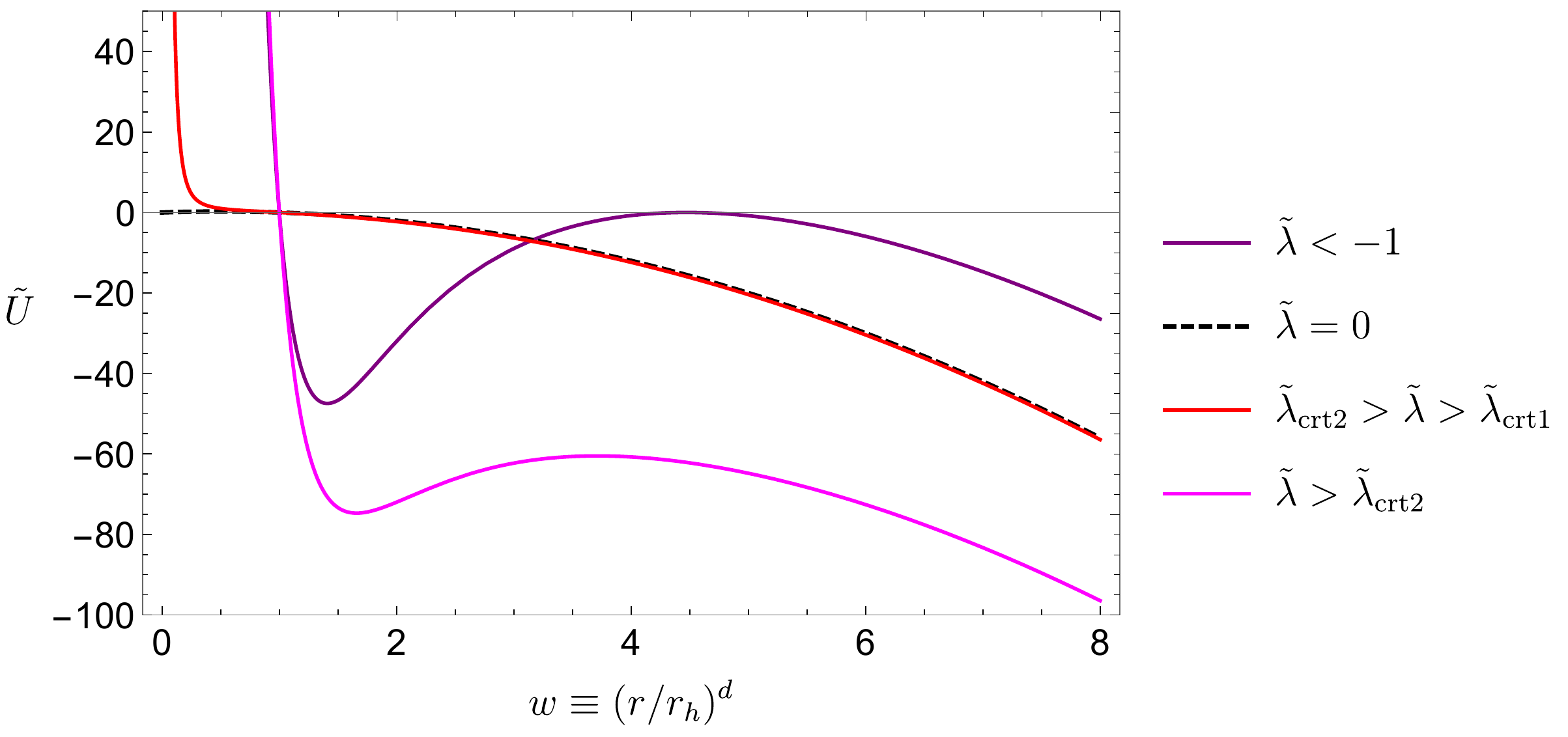}
	\caption{Left: The effective potentials $\widetilde{U}$ defined in eq.\eqref{eq:ham3} with $-1<\tilde{\lambda} < \tilde{\lambda}_{\rm{crt}1}$ present two extremal points inside the horizon which is located at $w=1$. Right: The effective potentials with $\tilde{\lambda} <-1$ or $\tilde{\lambda} > \tilde{\lambda}_{\rm{crt}1}$ have no extremal point inside the horizon. The dashed black curves in both plots denote the potential corresponding to the pure volume $V$.}
	\label{plot:potential}
\end{figure}
In order to explicitly illustrate the general discussion in the main text, let us consider a simple example of the generalized volume functional which is a covariant functional of the background curvatures evaluated on a codimension-one surface, \eg
\begin{equation}\label{eq:gv1}
\mathcal{C}_{\rm gen}= \frac{1}{\GN L} \int d^d\sigma \,\sqrt{h}\left(1 + \lambda \,L^4\,C^2 \right)\,, \quad \quad F_1=1 + \lambda \,L^4\,C^2 \,,
\end{equation}
where $h=\det h_{ij}$ is the determinant of the induced metric on the chosen hypersurface and $C^2\equiv C_{\mu\nu\rho\sigma}\,C^{\mu\nu\rho\sigma}$ denotes the square of the Weyl tensor for the bulk spacetime. Let us note that this $C^2$ term is the only nontrivial scalar functional that can be constructed from quadratic curvature terms. In particular, $\mathcal{R}^2$ and $\mathcal{R}^{\mu\nu}\mathcal{R}_{\mu\nu}$ simply reduce to constants when the bulk spacetime is a solution of the vacuum Einstein equations.
For the black hole background defined in eq.~\eqref{eq:vmetricBH}, one can explicitly obtain
\begin{equation}\label{eq:defineCC}
C^2\equiv \mathcal{R}_{\mu\nu\rho\sigma} \mathcal{R}^{\mu\nu\rho\sigma}-\frac{4}{d-1} \mathcal{R}_{\mu\nu} \mathcal{R}^{\mu\nu}+\frac{2}{d(d-1)} \mathcal{R}^{2}= d(d-1)^2(d-2)\,\frac{r_h^{2d}}{L^4 r^{2d}}\,.
\end{equation}
In this case, the function $a(r)$ appearing in the effective potential $\widetilde{U}(r)$ in e.q. \eqref{eq:Hamiltonian} is given by
\begin{equation}\label{eq:CCar}
a(r) = 1 + \tilde{\lambda}\,\left(\frac{r_h}{r}\right)^{2d}
\end{equation}
where we have redefined the dimensionless coupling as $\tilde{\lambda} =d(d-1)^2(d-2)\,\lambda$. The effective potential then reads
\begin{equation}\label{eq:ham3}
\begin{split}
{\widetilde U}(r)&=-\(\frac{r}{L}\)^{2(d-1)}\left(\frac{r^2}{L^2} - \frac{r_h^d}{L^2 r^{d-2}}\right)\,\left(
1 + \tilde{\lambda}\,\frac{r_h^{2d}}{r^{2d}}\right)^2 =\(\frac{r_h}{L}\)^{2d}\left(w-w^2\right)\left(
1 + \frac{\tilde{\lambda}}{w^2}\right)^2 \,,
\end{split}
\end{equation}
where we have defined a dimensionless radial coordinate $w\equiv (r/r_h)^d$. 
The extrema of the effective potential ${\widetilde U}(r)$ are determined by $\partial_r {\widetilde U}(r)=0 $, \ie 
\begin{equation}
\tilde{\lambda}= - w^2\,,\qquad \text{or} \qquad 
\tilde{\lambda} = \frac{w^2(2w-1)}{2w-3} \,.
\end{equation}
The first extremal solution, \ie $w=\sqrt{-\tilde{\lambda}}$ only exists for $\tilde{\lambda} <0$. The second extremal equation has two solutions when $\tilde{\lambda} > \tilde{\lambda}_{\mathrm{crt}2}= \frac{1}{8} (47+13\sqrt{13})$ or  $ 0\le \tilde{\lambda} < \tilde{\lambda}_{\mathrm{crt}1}= \frac{1}{8} (47-13\sqrt{13})$ but has no solution for ${\lambda}_{\mathrm{crt}1}<\tilde{\lambda} < {\lambda}_{\mathrm{crt}2}$. Figure \ref{plot:potential} shows various effective potentials with the Weyl-squared term and also the potential corresponding to the volume ($\tilde \lambda=0$). We further remark that the shape of the effective potential depends on neither the spacetime dimension nor the mass of the black hole. Both of these only appear in the overall coefficient in the effective potential. 

By examining the second derivative of the effective potential $\widetilde{U}''$ at the saddle points, it is easy to find that there are at most two extremal points and they correspond to a local minimum and a local maximum respectively. We will call the hypersurface which realizes the local maximum of the effective potential the final slice and we will denote it by $r=r_f$ or $w=w_f$. More explicitly, one finds
\begin{equation}\label{eq:finalSlice}
w_f=\frac{1}{6} \left(1+\frac{1+12 \tilde{\lambda} }{\({1-144 \tilde{\lambda} +6 \sqrt{3} \sqrt{-\tilde{\lambda}  (4 \tilde{\lambda}  (4 \tilde{\lambda} -47)+3)}} \)^{1/3}}+ \({1-144 \tilde{\lambda} +6 \sqrt{3} \sqrt{-\tilde{\lambda}  (4 \tilde{\lambda}  (4\tilde{\lambda} -47)+3)}} \)^{1/3}  \right) \,,
\end{equation}
with the corresponding value of the potential $\widetilde{U}(w_f) =  \frac{16 (1-w_f)^3 w_f}{(3-2 w_f)^2}\(\frac{r_h}{L}\)^{2d}$. Therefore, the local maximum of $\tilde U$ inside the horizon, corresponding to the late time surface, exists when $-1<\tilde \lambda<\tilde \lambda_{\rm crt1}$. In this case the effective potential has the qualitative form as shown in Fig. \ref{fig:potential}, and the general discussion in the main text applies. The cases with $\tilde \lambda<-1$ and $\tilde \lambda>\tilde \lambda_{\rm crt1}$ will be analyzed in \cite{longpaper}, where we show that they do not display a late time extremal surface.

Even in the case when the maximum in the potential exists, an important new feature compared to the volume ($\tilde \lambda=0$) is that $\tilde U$ diverges at the origin, therefore we have a new class of extremal surfaces, which come in from $r=\infty$, pass the maximum of the potential, and reflect off the infinite barrier near $r=0$. This means that at late times, the extremal surface is not unique. We will analyze this situation in more detail in \cite{longpaper} and show that the maximal generalized volume surface is the one that we analyzed in the present letter, \ie the one that approaches a constant radius surface, and does not make it past the maximum of the effective potential.


\section{B. Observables with $F_1\ne F_2$}\label{app:B}
We can also consider observables that are associated with the function $F_1$ but evaluated on the extremal surface determined by the functional $F_2$. Let's focus on the simplest case where $F_1, F_2$ both only depend on the bulk spacetime geometry. To wit, we define the generalized volume by 
\begin{equation}\label{eq:defineV12}
\begin{split}
O_{F_1,\Sigma_{F_2}} &= \frac{1}{\GN L} \int_{\Sigma_{F_2}} F_1(g_{\mu\nu}, \mathcal{R}_{\mu\nu\rho\sigma}, \nabla_\mu) \sqrt{h} \,d^d\sigma=\frac{V_x }{\GN L} \int_{\Sigma_{F_2}} d\sigma\,\(\frac{r}{L}\)^{d-1}\sqrt{-f(r\,){\dot v}^2+2\dot v\,\dot r}\ a_1(r)\,, 
\end{split}
\end{equation}
with the effective potential $ \widetilde{U}_1(r)= -f(r) \,a_1^2(r) \(\frac{r}{L}\)^{2(d-1)} $. The hypersurface where we evaluate the observable is the extremal surface associated with the potential $\widetilde{U}_2(r)= -f(r)\, a_2^2(r) \(\frac{r}{L}\)^{2(d-1)} $. The corresponding conserved momentum reads as 
\begin{equation}
\dot{r}^2+ \widetilde{U}_2(r) = P_v^2  \quad\text{with} \quad \widetilde{U}_2(r)= -f(r) a_2^2(r) \(\frac{r}{L}\)^{2(d-1)} \,.
\end{equation}
By using the extremization equations and our gauge choice, we can rewrite the generalized volume as follows
\begin{equation}
O_{F_1,\Sigma_{F_2}}=  -\frac{2V_x}{\GN L} \int^{r_{\rm{max}}}_{r_{\rm{min}}} dr \, \frac{ \sqrt{\widetilde{U}_1(r) \widetilde{U}_2(r)}}{f(r) \sqrt{P_v^2 - \widetilde{U}_2 (r)}}  \,.
\label{tort}
\end{equation} 
where $r_{\rm min}$ is determined by $\widetilde{U}_2(r_{\rm min})=P_v^2$ while $r_{\rm max}$ is some large cutoff radius which regulates the UV divergences associated with the observable. We note that this expression \reef{tort} reduces to the result for $F_1=F_2$ by setting $\widetilde{U}_2=\widetilde{U}_1$.

\subsection{Time Derivative and Linear Growth}
For simplicity, we denote the potentials at the minimal radius with $\widetilde{U}_i(r_{\rm{min}})= \bar{U}_i$. By using the integral for the boundary time in \eqref{eq:boundarytimetR02} with $\tilde U \rightarrow \tilde U_2$, one can decompose the observable as 
\begin{equation}
\frac{\GN L }{2 V_x}\, O_{F_1,\Sigma_{F_2}} = \sqrt{\frac{\bar{U}_1}{\bar{U}_2}}P_v t_\mt{R} +  \int^{r_{\rm{max}}}_{r_{\rm{min}}} \frac{dr}{f(r)} \frac{\sqrt{\bar{U}_1}P_v - \sqrt{\widetilde{U}_1(r)\widetilde{U}_2(r)}}{\sqrt{P_v^2 - \widetilde{U}_2 (r)}} \,.
\end{equation}
The time derivative of the observable at a general boundary time $\tau$ is thus given by
\begin{equation}
\frac{\GN L}{V_x }\frac{dO_{F_1,\Sigma_{F_2}}}{d\tau}=\sqrt{\bar{U}_1}+ 2P_v \frac{d P_v}{d \tau} \int^{\infty}_{r_{\rm{min}}} dr\, \frac{ \sqrt{\widetilde{U}_1(r)\widetilde{U}_2(r)} -\sqrt{\frac{\bar{U}_1}{\bar{U}_2}}\widetilde{U}_2(r) }{f(r) \(P_v^2-\widetilde{U}_2(r)\)^{3/2}}\,,
\end{equation}
where we have used $P_v^2 = \bar{U}_2$ to simplify the expression. Obviously, this time derivative includes not only a boundary term evaluated on $\Sigma_{\mt{CFT}}$ but also an integral term because now the hypersurface is determined by the extremality of $F_2$ rather than $F_1$. First of all, one can check that the growth rate is always finite, \ie 
\begin{equation}
\begin{split}
\frac{dO_{F_1,\Sigma_{F_2}}}{d\tau} &=\frac{V_x }{\GN L} \sqrt{\widetilde{U}_1(r_{\rm{min}})} + \mathcal{O}(1)\,,\\
\end{split}
\end{equation}
by performing the integral between $r_{\rm{min}}$ and the asymptotic boundary by using
\begin{equation}
\begin{split}
\sqrt{\frac{\bar{U}_1}{\bar{U}_2}}\widetilde{U}_2(r) - \sqrt{\widetilde{U}_1(r)\widetilde{U}_2(r)} \sim    \frac{\bar{U}_1 \widetilde{U}_2'(r_{\rm{min}})  -\bar{U}_2 \widetilde{U}_1'(r_{\rm{min}})}{2\sqrt{\bar{U}_1\bar{U}_2} }(r- r_{\rm{min}}) \,, \quad P_v^2 - \widetilde{U}_2(r)  \sim - (r- r_{\rm{min}})  \widetilde{U}_2'(r_{\rm{min}})  \,.
\end{split}
\end{equation}
It is also straightforward to find that the observable defined in eq.~\eqref{eq:defineV12} grows linearly in the late-time limit, namely
\begin{equation}
\begin{split}
\text{Linear Growth:}\qquad \lim_{\tau \to \infty} \frac{d O_{F_1,\Sigma_{F_2}}}{d\tau} & =\frac{V_x }{\GN L}\sqrt{\widetilde{U}_1(r_f)}\,.\\
\end{split}
\end{equation}
Of course, the linear growth is still connected to the fact that the effective potential associated with the extremal hypersurface has a local maximum at $\tau \to \infty, r_{\rm{min}}\to r_f$.  

Let us examine the next order corrections to the linear growth in the late-time limit, by analysing the second derivate of the observable.  Recalling the asymptotic behaviors 
\begin{equation}
\begin{split}
\lim_{\tau \to \infty}  \frac{d P_v}{d\tau}  &\sim (r_{\rm{min}} - r_f)^2 \,, \qquad \lim_{\tau \to \infty}   \frac{d r_{\rm{min}}}{d\tau}  \sim (r_{\rm{min}} - r_f) \,, \\
\lim_{\tau \to \infty}  \frac{d^2P_v}{d\tau^2}  &\sim (r_{\rm{min}} - r_f)^2 \,,\qquad \widetilde{U}_2(r) \approx  \widetilde{U}(r_f) +\frac{1}{2} \widetilde{U}''(r_f)(r-r_f)^2 \,,\\
\end{split}
\end{equation} 
we can obtain
\begin{equation}\label{eq:secondderivative01}
\lim_{\tau \to \infty} \frac{d^2 O_{F_1,\Sigma_{F_2}}}{d\tau^2} \sim \frac{V_x }{\GN L}\,\(r_{\rm{min}} - r_f\) \sim 0\,,\\
\end{equation} 
whose decay is interestingly slower than the case with $F_1=F_2$. Using the asymptotic expansion of $P_v(\tau)$ at late times, \ie 
\begin{equation}\label{eq:Pcrt}
P_{\infty}^2 - P_v^2 \sim -\frac{1}{2} \widetilde{U}''(r_f) \(r_{\rm{min}} -r_f \)^2\,,\qquad  P_{\infty} - P_v(\tau)   \sim   \# \,e^{- \kappa \tau}\,,
\end{equation}
we can show that the late-time limit of the growth rate of $O$ is given by 
\begin{equation}
\frac{dO_{F_1,\Sigma_{F_2}}}{d\tau}  \sim \frac{V_x }{\GN L}\,\(\sqrt{\widetilde{U}_1(r_f)} - \#' e^{-\kappa \tau/2}\) \,.\\
\end{equation}
This result is based on our general assumption that $r_f$ is {\it not} an extremum of the effective potential $\widetilde{U}_1$, \ie 
\begin{equation}
\widetilde{U}_1'(r_f) \ne 0\,,\qquad \frac{d\sqrt{\bar{U}_1}}{d \tau} \sim \frac{\widetilde{U}_1'(r_f) }{ 2 \sqrt{\widetilde{U}_1(r_f)}} \( r_{\rm{min}} - r_f  \) \,.
\end{equation} 
When $\widetilde{U}_1,\widetilde{U}_2$ share the same final radius $r_f$ as their maximal point ($\widetilde{U}_1'(r_f)=\widetilde{U}_2'(r_f)=0$), but not the value of the maximum $\widetilde{U}_1(r_f)\neq \widetilde{U}_2(r_f)$ one can instead get 
\begin{equation}\label{eq:secondderivative02}
\lim_{\tau \to \infty} \frac{d^2O_{F_1,\Sigma_{F_2}}}{d\tau^2} \sim
\frac{V_x }{\GN L}\,\(r_{\rm{min}} - r_f\)^2 \ln \( r_{\rm{min}} - r_f \) \sim 0\,.\\
\end{equation} 
Finally, in the case when both $\widetilde{U}_1'(r_f)=\widetilde{U}_2'(r_f)=0$ and $\widetilde{U}_1(r_f)=\widetilde{U}_2(r_f)$ we recover the behaviour of the case when $F_1=F_2$, \ie
\begin{equation}\label{eq:secondderivative03}
\lim_{\tau \to \infty} \frac{d^2O_{F_1,\Sigma_{F_2}}}{d\tau^2} \sim\frac{V_x }{\GN L}\, \(r_{\rm{min}} - r_f\)^2  \sim 0\,.\\
\end{equation}

\end{document}